\documentclass[twocolumn,showpacs,preprintnumbers,amsmath,amssymb]{revtex4}

\usepackage{amssymb}
\usepackage{amssymb}
\usepackage{epsfig,graphicx,times}
\usepackage{enumerate}
\usepackage{dcolumn}
\usepackage{color}
\usepackage{bm}

\newcommand{\beq}{\begin{equation}}
\newcommand{\eeq}{\end{equation}}

\newcommand{\bea}{\begin{eqnarray}}
\newcommand{\eea}{\end{eqnarray}}

\begin{document}


\title{Dynamics of coherences in the interacting double-dot Aharonov-Bohm interferometer:
Exact numerical simulations}

\author{Salil Bedkihal}
\affiliation{Chemical Physics Theory Group, Department of Chemistry, University
of Toronto, Toronto, Ontario M5S 3H6, Canada}

\author{Dvira Segal}
\affiliation{Chemical Physics Theory Group, Department of Chemistry, University
of Toronto, Toronto, Ontario M5S 3H6, Canada}

\begin{abstract}
We study the real time dynamics of electron coherence in a double
quantum dot two-terminal Aharonov-Bohm geometry, taking
into account repulsion effects between the dots' electrons.
The system is simulated by extending a numerically exact path
integral method, suitable for treating transport and dissipation in
biased impurity models [Phys. Rev. B 82, 205323 (2010)]. 
Numerical simulations at finite interaction strength are supported by master equation calculations in two
other limits: assuming
non-interacting electrons, and working in the Coulomb blockade regime.
Focusing on the intrinsic coherence dynamics between the double-dot states, we find that its temporal
characteristics are preserved under weak-to-intermediate inter-dot Coulomb interaction. 
In contrast, in the Coulomb blockade limit, 
a master equation calculation predicts coherence dynamics and a steady-state value which notably
deviate from the finite interaction case.
\end{abstract}

\pacs{
73.63.Kv,
03.65.Yz, 
73.63.-b, 
73.23.Hk  
}
 \maketitle


\section{Introduction}


Is electron transfer through quantum dot structures phase coherent,
or incoherent? How do electron-electron and electron phonon
interactions affect phase-coherent transport? From the other
direction, what is the role of the interference phenomena on
many-body effects, such as the formation of the Kondo resonance?
These questions were addressed in numerous experimental and
theoretical works, detecting the presence of quantum coherence in
mesoscale and nanoscale objects, using {\it Aharonov-Bohm} (AB)
interferometry, see for example Refs. \cite{Rev1, Heilb,2QDE,
Sigrist,GefenPRB,Gurvitz,Kubo, Col, Golosov, 
Kondo0,Boese, KondoHG,Kondo,Hod,Dhurba}.
In particular, oscillations  
in the conductance resonances of an AB interferometer, with either
one or two quantum dots embedded in its arms, were demonstrated in Refs. \cite{Heilb,2QDE},
indicating on the presence of quantum coherence.
Interestingly, AB oscillations were also manifested in the co-tunneling
regime, implying that phase coherence is involved within such processes \cite{Sigrist}.

Considering the role of electron-electron (e-e) interactions in the
AB interferometry, a systematic analysis carried out in Ref.
\cite{GefenPRB} has argued that spin flipping channels of the
transferred electron, the result of e-e repulsion effects, induce
dephasing. The consequence of this decohering effect was the
suppression of AB oscillations and the appearance of an
asymmetry in the resonance peaks.
One should note however that this study 
has assumed infinitely strong e-e interactions (Coulomb blockade regime) and treated the system 
perturbatively in the dot-metals coupling strength. In other studies, e-e repulsion
effects were totally ignored \cite{Konig-noint}, incorporated using a mean-field scheme, see for example
\cite{Golosov}, or treated perturbatively using the Green function formalism \cite{Green1,Green2}.
 These studies, and other theoretical and
numerical works \cite{GefenPRB,Ora}, have typically considered only the steady-state limit,
analyzing the conductance, a linear response
quantity, or the current behavior, often in the {\it infinite} large bias case \cite{Gurvitz,Kubo}.


The coherence of electron transfer processes through an AB
interferometer has been typically identified and
characterized via conductance oscillations in magnetic fields.
However, in a {\it double-dot} AB structure, a device including two
dots, both connected to biased metal leads, it is imperative that
the relative phase between the two dot states (or charge states)
should similarly convey information on electron coherence and
decoherence, as this phase is tangled with the AB phase. In a recent
work, Tu et al. \cite{Tu} have analyzed this intrinsic coherence
dynamics, revealing the effect of phase localization for 
different magnetic fluxes, by studying the real time dynamics of the two-dots reduced density
matrix. This analysis, based  on an exact (nonmarkovian) master
equation method \cite{Tulong}, left out e-e interaction effects all-together.

\begin{figure}[htbp]
\vspace{-12mm} \hspace{50mm} {\hbox{\epsfxsize=150mm
\epsffile{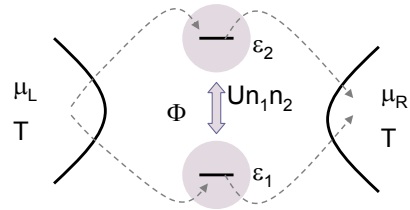}}}\vspace{-78mm} \caption{Scheme
of a double-dot AB interferometer. The two
dots are each represented by a single electronic level.
Electron repulsion energy is represented by the double arrow. The total
magnetic flux is denoted by $\Phi$.}  \label{FigS}
\end{figure}

Detailed study of the {\it dynamical} role of {\it finite}
electron-electron interactions on the {\it intrinsic} coherence
behavior in a {\it biased} double-dot AB interferometer, is the
focus of our work. 
The system includes a parallel quantum dot setup for
the AB interferometer, where (spinless) electrons experience an {\it
inter-dot} repulsion effect. For a schematic representation, see Fig.
\ref{FigS}. A unified description of the conductance behavior of
this model, a steady-state property, was given in Ref. \cite{Ora}.
Here, we focus on the dynamics of the
coherences, off diagonal elements of the double-dot reduced density
matrix. Furthermore, we simulate the charge current in the system, assuming different values for the
magnetic flux, at finite bias. Other effects considered are the role
of finite temperature on the coherence pattern, and the behavior
away from the electron-hole symmetric point, a regime not considered
before in a non-perturbative calculation within the AB setup \cite{Wacker}.

We follow the nonequilibrium real time dynamics of
electron coherence in this subsystem-bath model (double quantum
dot-metals) by performing exact numerical simulations, employing
the recently developed influence functional path integral (INFPI)
technique \cite{IF1,IF2}. This method relies on the observation
that in out-of-equilibrium (and/or finite temperature) cases bath
correlations have a finite range, allowing for their truncation
beyond a memory time dictated by the voltage-bias and the
temperature. Taking advantage of this fact, an
iterative-deterministic time-evolution scheme has been developed
where convergence with respect to the memory length can in principle
be reached. As convergence is facilitated at large bias, the method
is well suited for the description of the real-time dynamics of
mesoscale and nanoscale devices driven to a steady-state via interaction
with biased leads. The INFPI approach is complementary to other
numerically exact methods such as numerical renormalization group
techniques \cite{NRGBulla,Anders}, real time quantum Monte Carlo
simulations \cite{Phillip} and path integral methods \cite{egger}.
It offers flexibility in defining the impurity object and the metal
band structure. The results well converge at large voltage bias and/or
high temperatures, as we show below.

The principles of the INFPI approach have been detailed in Refs.
\cite{IF1,IF2}, where it has been adopted for investigating
dissipation effects in the nonequilibrium spin-fermion model, and the
population and the current dynamics in correlated
quantum dots, by investigating the single impurity Anderson model
\cite{Anderson} and the two-level spinless Anderson dot \cite{Sindel}.
In this paper, we further extend this approach, examining the effect of a
magnetic flux on the intrinsic coherence dynamics.
Our simulations show that general dynamical characteristics of the
double-dot coherence are maintained upon the application of
inter-dot Coulombic interactions. In particular, the characteristic
timescale for reaching the steady-state limit, the dependence of
the coherence on the AB phase factors, and the form of the temporal
current, similarly evolve for systems at zero or finite inter-dot 
interaction, for finite bias (beyond linear response), away from the
electron-hole symmetric point, at low or high temperatures.
We compare our data to (analytic) results based on a master equation treatment. This
method can readily handle the zero e-e interaction case and the counter case, the infinite interaction limit.
Interestingly, in the latter Coulomb blockade case the coherence is expected to evolve and sustain
values distinctively different from its behavior at finite interactions.

The paper is organized as follows. In Sec. II we describe the model
and draw the principles of the INFPI technique. In Sec. III we
present numerical results for the coherence dynamics and
the charge current, analyzing the role of electron-electron
interaction. Sec. IV includes analytic results based on 
master equations. Conclusions follow in Sec. V.


\section{Model and numerical method}


We focus on the symmetric AB setup, with a quantum dot
(impurity) located at each arm of the interferometer. The dots are
each connected to two metal leads (referred to as ``baths" or ``reservoirs"), maintained in
a biased state. For simplicity, we neglect the spin degree of
freedom and describe each quantum dot by a single spinless
electronic level. Overall, the dots '1' and '2' are represented by
the electronic levels $\epsilon_1$, and $\epsilon_2$, respectively,
described by the creation operators $d_m^{\dagger}$, $(m=1,2)$.
These levels are coupled in an AB geometry to two metal leads
$(\alpha=L,R)$ with chemical potentials $\mu_{\alpha}$.
For a schematic representation see Fig. \ref{FigS}.
The total Hamiltonian, $H$, includes the following terms
\bea
&&H= \epsilon_1 n_1 + \epsilon_2 n_2 + Un_1n_2 +
\sum_{\alpha,k}\epsilon_k  c_{\alpha,k}^{\dagger}c_{\alpha,k}
\nonumber\\
&&+ \sum_{k,m=1,2}\Big[V_{L,k,m}e^{i\phi_{m,L}}d_m^{\dagger}c_{L,k}
+ V_{R,k,m}e^{i\phi_{m,R}}
c_{R,k}^{\dagger} d_m
\nonumber\\
&&+h.c. \Big]
\label{eq:H}
\eea
Here, $c_{\alpha,k}^{\dagger}$ denotes the creation (annihilation)
of an electron with momentum $k$ in the $\alpha$ lead.
We assume identical leads, characterized by the same
band structure. For the subsystem,
$n_m=d_m^{\dagger}d_m$ represents the number operator for the impurity
level $m$, $U$ is the charging energy penalty for a simultaneous occupancy at the
two dots.
The AB phase factors, $\phi_{m,\alpha}$, are
acquired by electron waves under a magnetic field
perpendicular to the device plane,
\bea
\phi_{1,L}-\phi_{2,L}+\phi_{1,R}-\phi_{2,R}\equiv\phi=2\pi\Phi/\Phi_0.
\eea
Here $\Phi$ is the magnetic flux enclosed by the ring and
$\Phi_0=h/e$ is the flux quantum. In what follows, we adopt the
following gauge,
$\phi_{1,L}-\phi_{2,L}=\phi_{1,R}-\phi_{2,R}=\phi/2$. Besides the
phase factors, the coupling strength $V_{\alpha,k,m}$ are taken as
real numbers. The hybridization elements are given by
\bea
\Gamma_{\alpha,m,n}=\pi \sum_k V_{\alpha,k,m} V_{\alpha,k,n}e^{i(\phi_{m,\alpha}-\phi_{n,\alpha})}
\delta(\epsilon-\epsilon_{k}).
\eea
We assume that the couplings are identical for the two levels,
$V_{\alpha,k,m}=V_{\alpha,k,n}$, and
define the diagonal decay to the $\alpha$ bath 
\bea
\Gamma_{\alpha}=\pi\sum_{k}(V_{\alpha,k,m})^2
\delta(\epsilon-\epsilon_{k}).
\eea
The total diagonal decay is denoted by $\Gamma=\Gamma_L+\Gamma_R$.
In practice, we take $\Gamma_{\alpha}$ to be identical at the two ends.
Further, in what follows we only
consider the degenerate situation with $\epsilon\equiv \epsilon_m$. 

In the absence of magnetic fields this model is referred to as the
``spinless two-level Anderson model". It has been extensively
studied in the context of molecular electronics, for exploring
various effects in molecular conduction \cite{Nitzan}, and in
mesoscopic physics, revealing  nontrivial effects such as population
inversion \cite{Sindel, Ora} and transmission phase lapses
\cite{Meden, Karrasch, Wacker}.

Using the INFPI approach, the following observables could be followed:
the dots' occupation,
$\langle n_{m}\rangle \equiv {\rm Tr}[d_m^{\dagger}d_m \rho]$,
the coherence, $\sigma_{1,2}\equiv {\rm Tr}[
d_1^{\dagger}d_2 \rho]$, and the total current passing
through the interferometer. 
The trace is performed over all degrees of freedom, metals and impurity.
The charge current presented will be the symmetrized
current,  $\langle I_e\rangle \equiv {\rm Tr}[\hat I_e \rho]$,
accessed by defining the operator of interest as
\bea
\hat I_e&=&- \Im \sum_{k,m} V_{L,k,m}c_{L,k}^{\dagger}d_{m}e^{-i\phi_{m,L}}
\nonumber\\
&+&
\Im \sum_{k,m} V_{R,k,m}c_{R,k}^{\dagger}d_{m}e^{i\phi_{m,R}},
\eea
with $\Im$ denoting the imaginary part. Within INFPI, these observables are
simulated in the Heisenberg representation as we explain below,
assuming an initial density matrix $\rho(0)$ describing a
nonequilibrium-biased situation.



We outline now the principles of the INFPI method,
allowing for the exact simulation of transport and
dissipation in impurity models \cite{IF1,IF2}.
We begin by reorganizing the Hamiltonian, Eq. (\ref{eq:H}), as $H=H_0+H_1$,
identifying the nontrivial many-body interaction term as
\bea
H_1= U\left[n_1n_2-\frac{1}{2}(n_1+n_2)\right].
\eea
%
$H_0$ contains the remaining two-body terms, redefining the dot energies
as $E_{d,m}=\epsilon_m+U/2$. This partitioning allows us to utilize
the Hubbard-Stratonovich (HS) transformation \cite{Hubb-Strat}, see
Eq. (\ref{eq:HS}) below. Formally, the dynamics of a quadratic operator, $\hat
A$, either given in terms of the baths (metals) or impurity degrees
of freedom, can be written as
\bea \langle \hat A(t) \rangle = {\rm Tr} [\rho(0) \hat A(t)] =
\lim_{\lambda \rightarrow 0} \frac{\partial}{\partial \lambda} {\rm
Tr}\big[\rho(0) e^{iHt}e^{\lambda \hat A}e^{-iHt} \big].
\label{eq:At} \eea
Here $\lambda$ a real number, taken to vanish at the end of the calculation, 
 $\rho$ is the total density matrix, and the trace is performed
over both subsystem and reservoirs degrees of freedom. For simplicity, we
assume that at the initial time $t=0$ the dots and the baths are
decoupled,
$\rho(0)=\sigma(0) \otimes \rho_L\otimes \rho_R$.
The baths are prepared in a nonequilibrium
biased state  $\rho_{\alpha}$; the subsystem is
described by the (reduced) density matrix $\sigma(0)$.

We proceed and factorize the time evolution operator, $e^{iHt} =
(e^{iH\delta t})^N$, further utilizing the Trotter decomposition
$e^{iH\delta t}\approx\big( e^{iH_0\delta t/2} e^{iH_1 \delta t}
e^{iH_0\delta t/2}  \big)$. The many-body term $H_1$ can be
eliminated by introducing auxiliary Ising variables $s=\pm$ via the
HS transformation \cite{Hubb-Strat},
\bea
e^{\pm iH_1 \delta t} &=& \frac{1}{2} \sum_{s}
e^{H_\pm(s)}; \,\,\,\,
e^{H_\pm(s)}\equiv
e^{-s \kappa_{\pm} (n_{2}-n_{1})}.
\label{eq:HS}
\eea
Here $\kappa_{\pm}=\kappa' \mp i \kappa'' $,
$\kappa'=\sinh^{-1}[\sin(\delta t U/2)]^{1/2}$, $\kappa''
=\sin^{-1}[\sin (\delta t U/2)]^{1/2}$.
The uniqueness of this transformation requires $U \delta t < \pi$.
Incorporating the Trotter decomposition and the HS transformation into Eq. (\ref{eq:At}),
the time evolution of $\hat A$ is dictated by
\bea
\langle \hat A(t)\rangle =
\lim_{\lambda \rightarrow 0} \frac{\partial}{\partial \lambda} \Big \{
\int ds_1^{\pm} ds_2^{\pm},..., ds_N^{\pm}
I(s_{1}^{\pm}, s_2^{\pm},..., s_N^{\pm})
\Big \}.
\label{eq:At2}
\eea
The integrand, referred to as as the ``Influence Functional" (IF), is given by
($k=1$, $k+p=N$)
\begin{widetext}
\bea
 I(s_k^{\pm},..., s_{k+p}^{\pm})=
\frac{1}{2^{2(p+1) }}
{\rm Tr} \Big[\rho(0)
\mathcal
G_+(s_{k+p}^+) ... \mathcal G_+(s_{k}^+) e^{i H_0 (k-1) \delta
t} e^{\lambda  {\hat A}} e^{-i H_0 (k-1) \delta t} \mathcal
G_-(s_{k}^-)... \mathcal G_-(s_{k+p}^-)  \Big].
\nonumber\\
\label{eq:IF}
\eea
\end{widetext}
Here $\mathcal G_{+}(s_k^{+}) = \left( e^{i H_0 \delta t/2} e^{
H_{+}(s_{k}^{+})}  e^{i H_0 \delta t/2} \right)$ and $\mathcal
G_-=\mathcal G_+^{\dagger}$. Eq. (\ref{eq:At2}) is exact
in the $\delta t\rightarrow 0$ limit. 
Practically, it can be evaluated by noting that
in standard nonequilibrium situations, even at zero
temperature, bath correlations die exponentially, thus the IF in Eq.
(\ref{eq:At2}) can be truncated beyond a memory time
$\tau_c=N_s\delta t$, corresponding to the time beyond which bath
correlations may be controllably ignored \cite{IF1}. Here $N_s$ is an integer
and the correlation time $\tau_c$ is determined by the nonequilibrium
situation, roughly $\tau_c \sim 1/\Delta \mu$. This argument implies
the following (non-unique) breakup  \cite{IF1}
\bea
&&I(s_{1}^{\pm}, s_{2}^{\pm},...s_N^{\pm} )
\simeq
I(s_1^{\pm},s_2^{\pm},..., s_{N_s}^{\pm})
I_s(s_2^{\pm},s_3^{\pm},..., s_{N_s+1}^{\pm}) ...
\nonumber\\
&& \times
I_s(s_{N-N_s+1}^{\pm},s_{N-N_s+2}^{\pm},..., s_{N}^{\pm}),
\label{eq:dynamics}
\eea
where each element in the product,
besides the first one, is given by a ratio between truncated IF,
\bea I_s(s_k,s_{k+1},...,s_{k+N_s-1})=
\frac{I(s_k^{\pm},s_{k+1}^{\pm},...,s_{k+N_s-1}^{\pm})}{I(s_{k}^{\pm},s_{k+1}^{\pm},...,s_{k+N_s-2}^{\pm})}.
\label{eq:Is}
\eea
%
It is useful to define the multi-time object
\bea
&&{\mathcal R}(s_{k+1}^{\pm}, s_{k+2}^{\pm},..., s_{k+N_s-1}^{\pm} ) \equiv
\nonumber\\
&&\sum_{s_1^{\pm},s_2^{\pm},..., s_{k}^{\pm}} I(s_1^{\pm},s_2^{\pm},..., s_{N_s}^{\pm})
I_s(s_2^{\pm},s_3^{\pm},..., s_{N_s+1}^{\pm})...
\nonumber\\
&&\times  I_s(s_{k}^{\pm},s_{k+1}^{\pm},..., s_{k+N_s-1}^{\pm}),
\label{eq:R1}
\eea
and time-evolve it by multiplying it with the subsequent truncated IF, then summing
over the intermediate variables,
\bea
&&{\mathcal R}(s_{k+2}^{\pm}, s_{k+3}^{\pm},..., s_{k+N_s}^{\pm} )=
\nonumber\\
&&\sum_{s_{k+1}^{\pm}} {\mathcal R}(s_{k+1}^{\pm}, s_{k+2}^{\pm},..., s_{k+N_s-1}^{\pm} )
I_s(s_{k+1}^{\pm},s_{k+2}^{\pm},...,s_{k+N_s}^{\pm}).
\label{eq:R2}
\nonumber\\
\eea
Summation over the internal variables results in the {\it time local} expectation value,
\bea
\langle e^{\lambda \hat A(t_k)} \rangle =
\sum_{s_{k+2-N_s}^{\pm},...,s_{k}^{\pm}}
{\mathcal R}(s_{k+2-N_s}^{\pm}, s_{k+3-N_s}^{\pm},..., s_{k}^{\pm} ).
\label{eq:R3}
\eea
This procedure should be repeated for several (small) values of
$\lambda$. Taking the numerical derivative with respect to
$\lambda$, the expectation value $\langle \hat A(t_k)\rangle$  is
retrieved.

The main element in this procedure, the truncated IF [Eq. (\ref{eq:IF})],
is calculated using a fermionic trace formula \cite{Klich},
\bea
I&=&{\rm Tr} \left[e ^{M_1} e^{M_2}...e ^{M_p} (\rho_L \otimes
\rho_R \otimes \sigma(0)) \right]
\nonumber\\
&=& {\rm det} \Big \{  [I_L-f_L] \otimes  [I_R-f_R]  \otimes  [I_S-f_S].
\nonumber\\
&+& e^{m_1}e^{m_2}...e ^{m_p} [ f_L \otimes f_R \otimes  f_S] \Big
\}.
\label{eq:IT}
\eea
Here, $\rho_{\alpha}$, the time-zero density matrix of the
$\alpha=L,R$ fermion bath and $\sigma(0)$, the subsystem initial density matrix,
are assumed to follow an exponential form. Other terms $e^{M}$, with $M$ a quadratic
operator, represent further factors in Eq. (\ref{eq:IF}).
In the determinant,
$m$ is a single-particle operator, corresponding to the quadratic operator 
$M=\sum_{i,j}(m)_{i,j}c_i^{\dagger}c_j$; $c_i^{\dagger}$ ($c_j$) are fermionic 
creation and annihilation operators, either related to the system or the baths. 
The matrices $I_{\alpha}$ and $I_S$ are the
identity matrices for the $\alpha$ space and for the subsystem,
respectively. The functions $f_L$ and $f_R$ are the bands electrons'
energy distribution,
$f_{\alpha}=[e^{\beta_{\alpha}(\epsilon-\mu_{\alpha})}+1]^{-1}$,
with the chemical potential $\mu_{\alpha}$ and inverse temperature
$\beta_{\alpha}$.

The determinant in Eq. (\ref{eq:IT}) is evaluated numerically by
taking into account $L_s$ electronic states for each metal. This
discretization implies a numerical error. However, we have found
that with $L_s\sim 100$ states we can reach convergence in the time
interval of interest.
Other sources of error, elaborated and examined in
Refs. \cite{IF1, IF2}, are the {\it Trotter error}, originating
from the approximate factorization of the total Hamiltonian into the
non-commuting $H_0$ (two-body) and $H_1$ (many-body) terms, and the
{\it memory error}, resulting from the truncation of the IF.
Convergence is verified by demonstrating
that results are insensitive to the time step and the memory size,
once the proper memory time is accounted for.

As we show below, distinct observables may require different
memory time $\tau_c$ for reaching convergence: The dots' occupation
and the {\it real part} of the subsystem off-diagonal element,
$\Re\sigma_{1,2}$, converge for $\tau_c\sim 1/\Delta\mu$. In
contrast, the charge current and $\Im \sigma_{1,2}$ require a memory time at least twice longer,
as these quantities are sensitive to the
bias drop at each contact, rather than to the overall voltage bias.
It is important to note that this scaling
is approximate, and the actual memory time
further depends on the subsystem (dots) energetics in a complex way:
First, the memory time depends on $U$ in a nontrivial manner
\cite{RabaniGuy}. In the absence of $U$  INFPI numerical results
are exact, irrespective of the memory size used in the simulation.
This can be seen from Eqs. (\ref{eq:IF}) and (\ref{eq:dynamics}),
where a cancellation effect takes place leaving free propagation
terms only, from $t=0$ to the current time. At infinitely large $U$
one expects again superior convergence behavior, as simultaneous occupancy
is forbidden \cite{RabaniGuy}. Second, the position of the dot
states with respect to the left and right chemical potentials affect
the convergence behavior. We generally found that when the
dot states are located {\it within} the bias window a shorter memory
time is required for reaching convergence, in comparison to the case
where the dot energies are out-of-resonance with the bias window.
This could be rationalized by noting that the decorrelation time for
electrons within the bias window is short relative to the
characteristic timescale of electrons occupying off-resonance states.



\section{INFPI numerical results}

We present here data for the coherence dynamics $\sigma_{1,2}(t)=\langle
d_1^{\dagger}(t)d_2(t) \rangle$ and the charge current $\langle I_e\rangle$ within the
interacting double-dot AB interferometer.
As we show below, we find that finite e-e interactions do not destroy the general characteristics of the
coherence behavior, for the cases $U/\Gamma\leq 4$ considered here.
We focus on the following set of parameters: The double-dot
subsystem includes two degenerate states with
$\epsilon\equiv\epsilon_m$ ($m=1,2$). The dynamics is studied away
from the electron-hole symmetric point, $E_d=\epsilon+U/2=0.2$. The
metals' band structure is taken identical at the two ends, and we
use leads with constant density of states and a sharp cutoff at
$D=\pm 1$. The inter-dot repulsion is taken at the range $U=0-0.2$, whereas the
system-bath hybridization strength (see definitions in Sec. II)
 is taken as $\Gamma=0.05$. 
As we demonstrate below, our results generally converge for
 $U/\Gamma\leq4$.
The bias voltage is applied in a symmetric manner, $\mu_L=-\mu_R$, and we take
$\mu_L-\mu_R=\Delta\mu \sim 0.6$. The temperature is varied, where $\beta=1/T=200$ 
corresponds to the low-$T$ case, and $\beta=5$ reflects a high-$T$ situation.
The numerical parameters adopted are $L_s\sim 100$ states per bath,
time step of $\delta t\sim 0.8-1.6$ and a memory time $\tau_c\sim
3-10$. This  choice of bath states suffices for mimicking a
continuous band structure \cite{IF1,IF2}. Also, recurrence effects
are not observed before $t\Gamma\sim 10$. For simulating dynamics
beyond that time larger reservoirs are constructed, as necessary.
The time step was selected based on two (contrary)
considerations: (i) It should be made short enough, for justifying
the Trotter breakup, $\delta t U<1$. (ii) For computational reasons, it should be made long
enough, to allow coverage of the system memory time with few
terms, $N_s<8$, recalling that $\tau_c=\delta t N_s$.

Before presenting our results we explain the initial condition
adopted. At time $t=0$ the double-dot levels are both empty, while
the (decoupled) reservoirs are separately prepared with occupation functions
obeying the Fermi-Dirac statistics at a given temperature $T$
and bias.


\begin{figure}[htbp]
\vspace{0mm} \hspace{0mm} {\hbox{\epsfxsize=65mm \epsffile{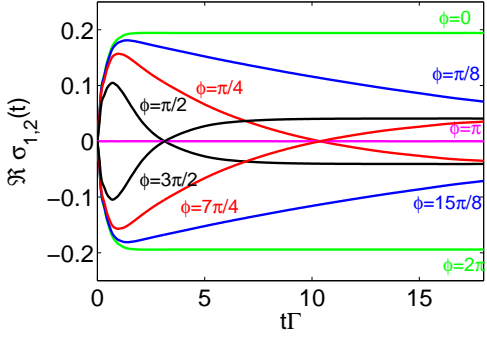}}}
\caption{Time evolution of the states coherence, in the absence of
electron repulsion effects. Shown is the real part of
$\sigma_{1,2}(t)$, plotted for the phases $\phi$ ranging from 0 to
$2\pi$, top to bottom. $E_d=0.2$, $\Gamma=0.05$, $U=0$,
$\Delta\mu=0.6$, $\beta=200$, $L_s=240$.} \label{Fig1}
\end{figure}

\begin{figure}[htbp]
\vspace{0mm} \hspace{0mm} {\hbox{\epsfxsize=67mm
\epsffile{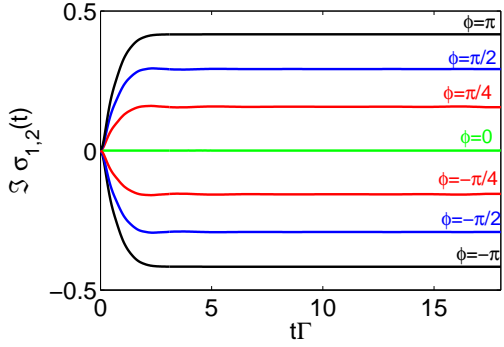}}} \caption{
Time evolution of the imaginary part of $\sigma_{1,2}(t)$,
in the absence of electron repulsion effects.
The phase factors $\phi$ range between $-\pi$ to $\pi$, bottom to top.
Other parameters are the same as in Fig. \ref{Fig1}.
}
\label{Fig2}
\end{figure}

\subsection{Coherence dynamics at $U=0$}

We begin by presenting results for the noninteracting case,
$U=0$. Figures \ref{Fig1} and \ref{Fig2} display the time evolution
of the real and imaginary parts of $\sigma_{1,2}(t)$, respectively,
for relatively large bias $\Delta \mu=0.6$ and at low temperature. We find that
$\Re \sigma_{1,2}$ decays at a flux dependent rate
 after the initial rise. The imaginary part, displayed in Fig. \ref{Fig2},
saturates with a time scale $\Gamma$ \cite{Tu}.
Defining $\sigma_{1,2}(t)=|\sigma_{1,2}(t)|e^{i\varphi(t)}$, it was
argued in Ref. \cite{Tu} that this relative phase localizes to the values
$\varphi=-\pi/2$ or $\pi/2$ in the long time limit when $\phi\neq
2p\pi$, $p$ is an integer. This localization
behavior is expected only when the (degenerate) dot levels are
{\it symmetrically} placed between the chemical potentials, i.e., for
$\epsilon=0$.
Away from this symmetric point, using $\epsilon=0.2$, Fig.
\ref{Fig1}  shows that the real part of $\sigma_{1,2}$ is finite
in the asymptotic limit for any phase, besides $\pi$. It is
interesting to note though that when $\phi\neq 2p\pi$, $\Re
\sigma_{1,2}$ still approaches a certain-fixed value, irrespective of the
magnetic flux.


\begin{figure}[htbp]
\vspace{0mm} \hspace{0mm} {\hbox{\epsfxsize=70mm \epsffile{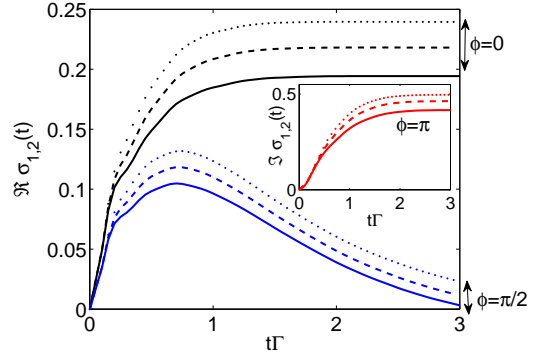}}}
\caption{Time evolution of  $\sigma_{1,2}$ for $U=0$ (full line), $U=0.1$
(dashed line) and $U=0.2$ (dotted line). Main: Real part of
$\sigma_{1,2}(t)$. The three top lines were simulated for $\phi=0$.
The bottom lines were obtained using $\phi=\pi/2$. The numerical
parameters are $\delta t=1$, $N_s=6$ and $L_s=120$. Inset:
Imaginary part of $\sigma_{1,2}(t)$ when $\phi=\pi$. Numerical
parameters are $\delta t=1.6$, $N_s=6$ and $L_s=120$. Other parameters
are $E_d=0.2$, $\Gamma=0.05$, $\Delta\mu=0.6$ and $\beta=200$.} \label{Fig3}
\end{figure}

\begin{figure}[htbp]
\vspace{0mm} \hspace{0mm} {\hbox{\epsfxsize=70mm
\epsffile{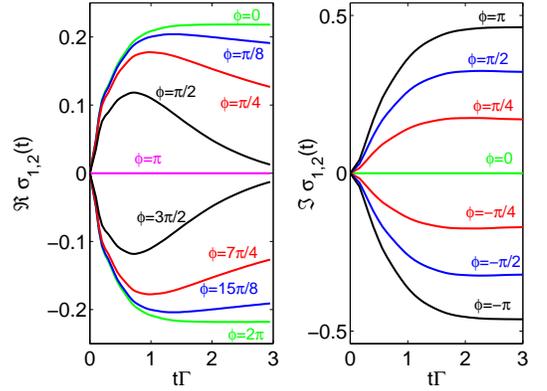}}}
\caption{Time evolution of the states coherence for $U=0.1$, for several phase factors.
$E_d=0.2$, $\Gamma=0.05$, $\Delta\mu=0.6$, $\beta=200$.
The real part of $\sigma_{1,2}$ was obtained with $\delta t=1$ and $N_s=6$;
the imaginary part was simulated with $\delta t=1.6$ and $N_s=6$.
}
\label{Fig4}
\end{figure}

\subsection{Coherence and current at finite $U$}

We now investigate the role of e-e repulsion effects on the
coherence behavior. Fig. \ref{Fig3} displays the real part of
$\sigma_{1,2}(t)$ for two phases, $\phi=0$ and $\phi=\pi/2$, and its
imaginary part for $\phi=\pi$ (inset), for three values of $U$.
Data for $\Im\sigma_{1,2}(t)$ at $U=0.2$ has not yet converged for the $\tau_c$ adopted,
see text following Fig. \ref{Figc2}.
In comparison to the $U=0$ case, we find that general trends are
maintained, though the long time coherences are larger in the finite
$U$ case. Note our convention: the parameter $E_d=\epsilon+U/2$ is
maintained fixed between simulations with different values of $U$. The
trajectory simulated extends up to $\Gamma t=3$, where convergence
is satisfactory. Different memory times were used
for simulating the real part of the coherence, and its imaginary part.
We adopted $\tau_c\sim 5$ when simulating
$\Re\sigma_{1,2}$,  whereas $\tau_c\sim 10$ was used for acquiring  $\Im \sigma_{1,2}$. A more
detailed discussion of convergence issues is given in Sec. III.C.

\begin{figure}[htbp]
\vspace{0mm} \hspace{0mm} {\hbox{\epsfxsize=65mm \epsffile{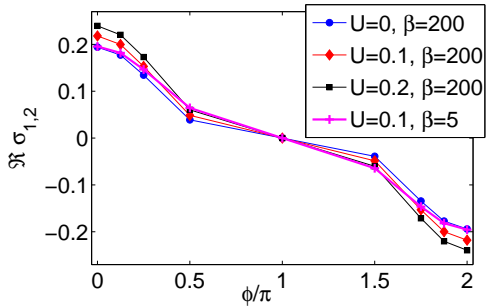}}}
\caption{Effect of finite $U$ on the coherence.
$\Re \sigma_{1,2}$ is plotted as a function of the phase factor $\phi$ at a particular time,
$\Gamma t=2$, for different $U$-values and temperatures.
Other parameters are the same as in Fig. \ref{Fig4}.}
\label{Fig5}
\end{figure}

\begin{figure}[htbp]
\vspace{0mm} \hspace{0mm} {\hbox{\epsfxsize=65mm \epsffile{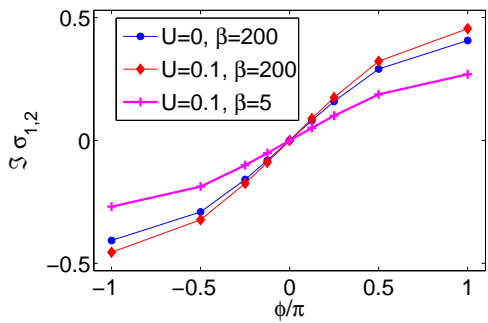}}}
\caption{Effect of finite $U$ on the coherence.
$\Im\sigma_{1,2}$ is plotted as a function of $\phi$ at a particular time,
$\Gamma t=2$, at finite $U$ and for different temperatures.
Other parameters are the same as in Fig. \ref{Fig4}.}
\label{Fig5a}
\end{figure}

Fig. \ref{Fig4} presents 
$\sigma_{1,2}(t)$ for several phases $\phi$, at $U=0.1$. By comparing the data to the
zero-$U$ case (see Figs. \ref{Fig1} and \ref{Fig2}), we conclude that the symmetry of the off-diagonal
elements is maintained in the presence of $U$.
The general pattern of the coherence is displayed in
Figs. \ref{Fig5} and \ref{Fig5a}, plotting the behavior
of $\sigma_{1,2}$ as a function of the phase factor, at a particular time,
$\Gamma t=2$, for $U=0$, 0.1, and 0.2, at different temperatures.
It should be noted that by this time the real part of
the coherence has {\it not} yet reached its steady-state value.
%
We find that the coherence symmetry around $\phi=\pi$ (for $\Re
\sigma_{1,2}$) or $\phi=0$ (for $\Im \sigma_{1,2}$) is maintained,
though the absolute numbers change. Interestingly, while the effect
of the temperature is significant for $\Im \sigma_{1,2}$,
showing visible reduction in values at high $T$, the real part
of $\sigma_{1,2}$ is only lightly affected by the temperature. The
downfall of $\Im \sigma_{1,2}$ with temperature is also reflected in
the behavior of the charge current, as we show next.

\begin{figure}[htbp]
\vspace{0mm} \hspace{0mm} {\hbox{\epsfxsize=75mm
\epsffile{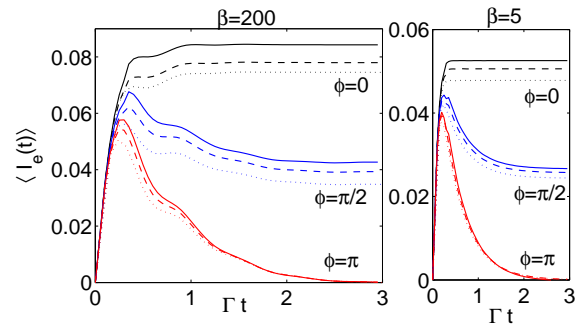}}}
\caption{
Charge current through an AB interferometer at low temperatures, $\beta=200$ (left panel)
and high temperatures $\beta=5$ (right panel)
for $\phi=0$, $\pi/2$, and $\pi$, top to bottom with
$U=0.2$ (full line), $U=0.1$ (dashed line), $U=0$ (dotted line).
Other parameters are the same as in Fig. \ref{Fig4}.
Numerical parameters are $\delta t=1$, $N_s$=6 and $L_s$=120.
}
\label{FigI1}
\end{figure}

We study the behavior of the charge current at different phases, for different
e-e repulsion strengths and temperatures. 
Fig. \ref{FigI1} shows, as expected, a
destructive interference pattern for electron current in the long
time limit when $\phi=\pi$, irrespective of the value of $U$. This
perfect destructive interference indicates that charge transport is
fully coherent in this model. The temporal behavior does show
however a sensitivity to the value of $U$,  manifesting that systems
with variable $U$ differently respond to the initial condition. 

In the steady-stat limit the current scales like $\langle I_e\rangle
\propto [1+\cos(\phi)]$, for finite $U$
\cite{GefenPRB}. This relation does not hold in the short time limit.
It is interesting to note that irrespective of $U$
and the phase factor the current approaches the steady-state limit on a
relatively short timescale, $\Gamma t\sim 2$. At high temperatures,
Fig. \ref{FigI1} manifests that the system is still fully coherent,
while temporal oscillations are washed out. The reduction of the
current at high temperatures can be attributed to the softening of
the contacts' Fermi functions from the sharp step-like form at
low temperatures. The electronic states at the right lead in the bias window are
not fully empty any longer.
Similarly, at the left lead electronic states overlapping with $E_d$ may be empty.
Overall, this  results in the reduction of the current at high $T$.



\subsection{Convergence analysis}

\begin{figure}[htbp]
\vspace{0mm} \hspace{0mm} {\hbox{\epsfxsize=70mm \epsffile{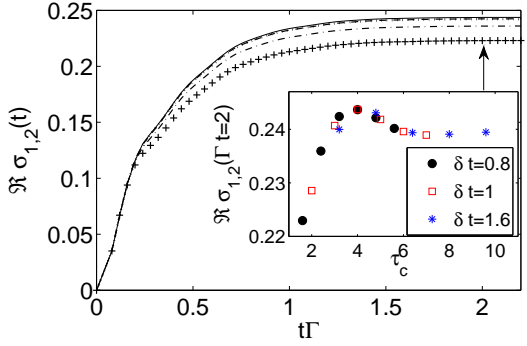}}}
\caption{Convergence behavior of $\Re \sigma_{1,2}$ for $\phi=0$ and
$U$=0.2. Other physical parameters are the same as in Fig. \ref{Fig4}.
Numerical parameters are $\delta t=0.8$ and $N_s=2$ ($+$), $N_s=3$
(dashed-dotted line), $N_s=4$ (dashed line), $N_s=5$ (full line) and $N_s=6$
(dotted line). The inset zooms on the convergence 
 at a particular time, $\Gamma t=2$, as a function of the  memory time
$\tau_c=N_s\delta t$, using three different values for the time steps, $\delta
t=0.8$ ($\circ$), $\delta t=1$ ($\square$) $\delta t=1.6$  ($*$). }
\label{Figc1}
\end{figure}

We exemplify here the convergence behavior of the real and imaginary parts of $\sigma_{1,2}$ at
low temperatures, as well as the behavior of the current.
Fig. \ref{Figc1} demonstrates that $\Re\sigma_{1,2}$ nicely converges for
$U=0.2$, for $\tau_c\geq 5$. The asymptotic limit is
practically reached, within $\sim 1.5\%$ error, already for
$\tau_c\sim1/\Delta \mu$. We confirm that the results are
insensitive to the particular time step selected (inset).
We have also verified (not shown) that simulations performed with different
phase factors similarly converge.

\begin{figure}[htbp]
\vspace{0mm} \hspace{0mm} {\hbox{\epsfxsize=75mm \epsffile{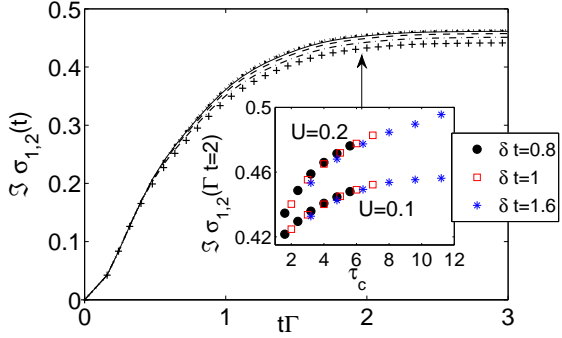}}}
\caption{Convergence behavior of $\Im \sigma_{1,2}$ for $\phi=\pi$ and
$U$=0.1. Other physical parameters are the same as in Fig. \ref{Fig4}.
Numerical parameters are $\delta t=1.6$ and $N_s=2$ ($+$), $N_s=3$
(dashed-dotted), $N_s=4$ (dashed line), $N_s=5$ (full line) and $N_s=6$ 
(dot), $N_s=7$ (dotted line). The inset presents $\Im
\sigma_{1,2}$ at  a particular time,  $\Gamma t=2$, for $U=0.1$ and $U=0.2$, as a function of
the memory time $\tau_c=N_s\delta t$, using three different time
steps, $\delta t=0.8$ ($\circ$), $\delta t=1$ ($\square$) $\delta
t=1.6$  ($*$).} \label{Figc2}
\end{figure}

The convergence of $\Im \sigma_{1,2}$ is generally 
slower, as we show in Fig. \ref{Figc2}. While $\Re \sigma_{1,2}$
converges for $\tau_c \gtrsim 1/\Delta \mu$, we find that $\Im
\sigma_{1,2}$ requires memory time at least twice longer for
achieving convergence. For $U=0.1$ $\Im\sigma_{1,2}$ is converging.
In contrast, at stronger interactions, $U=0.2$, the
large time step adopted results in a Trotter error buildup, and
the results seem to diverge around $\tau_c \sim 10-12$ (inset). 

\begin{figure}[htbp]
\vspace{0mm} \hspace{0mm} {\hbox{\epsfxsize=70mm \epsffile{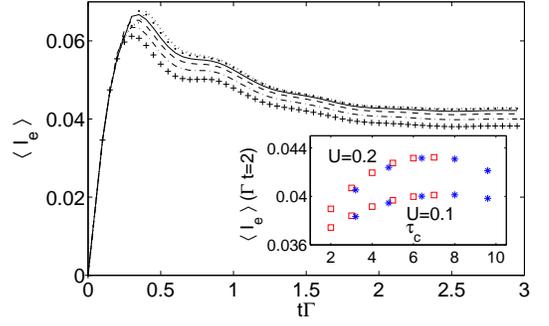}}}
\caption{Convergence behavior of the charge current, $\phi=\pi/2$ and
$U$=0.2. Other physical parameters are the same as in Fig. \ref{Fig4}.
Numerical parameters are $\delta t=1$ and $N_s=2$ ($+$), $N_s=3$
(dashed-dotted), $N_s=4$ (dashed line), $N_s=5$ (full line) and $N_s=6$
(dot), $N_s=7$ (dotted line). The inset presents the data
 at a particular time,  $\Gamma t=2$, for $U=0.1$ (bottom) and $U=0.2$ (top), as a function of
the memory time $\tau_c=N_s\delta t$, using 
$\delta t=1$ ($\square$) and $\delta t=1.6$  ($*$).} \label{Figc3}
\end{figure}

We also present the behavior of the charge current at different $\tau_c$ values,
see Fig. \ref{Figc3}. It generally converges when $\tau_c\sim 6$,
irrespective of the phase factor (not shown), for $U/\Gamma \leq 4$, in agreement with
earlier studies \cite{IF2}.

Overall, we conclude that we can faithfully simulate the time evolution of the
coherence $\sigma_{1,2}$ and the current for $\Delta\mu=0.6$ and $U/\Gamma=2$. For larger $U$,
the real part of $\sigma_{1,2}$,  the dot occupation, and the current can be
still converged \cite{IF1,IF2}. The simulation of $\Im \sigma_{1,2}$
requires longer $\tau_c$ and a shorter time step at $U/\Gamma>2$.
Roughly, these observations can be rationalized noting that the dynamics of $\Re
\sigma_{1,2}$ is influenced by the {\it full potential drop},
$\mu_L-\mu_R$, similarly to the dots occupation $\langle n_{m}\rangle$ \cite{Tu}.
In contrast, the dynamics of $\Im
\sigma_{1,2}$ is sensitive to the bias drop at {\it each contact} \cite{Tu}, 
resulting in longer decorrelation times.


\section{Master equation analysis:  $U=0$ and $U=\infty$}

Rate equations for resonant transport in interacting multidot structures can be derived 
based on the microscopic many-body Schr\"odinger equation  \cite{Gurvitz98}.
We support INFPI numerical simulations with an analytical study of the
system's dynamics, based on such a master equation description. Specifically, 
we adopt the Bloch-type equations derived in Ref. \cite{Gurvitz}, for the
reduced density matrix of the double-dot system in the charge state
basis $\sigma_{j,j'}(t)$, $j={a,b,c,d}$. Here the index $j$ labels
the double-dot charge states in order of an empty dot ($a$), single
occupied dot, on either the "1" or "2" sites ($b$ and $c$ states,
respectively) and the state ($d$), with the two dots occupied.
Explicitly, $|a\rangle \leftrightarrow |0 0\rangle$, $|b\rangle
\leftrightarrow |1 0\rangle$, $|c\rangle \leftrightarrow |0
1\rangle$, and $|d\rangle \leftrightarrow |1 1\rangle$. The creation
and annihilation operators of the dot are related to this states by
$d_{1}^{\dagger}\leftrightarrow|00\rangle\langle 0 1|+
|01\rangle\langle 1 1|$ and
 $d_{2}^{\dagger}\leftrightarrow|00\rangle\langle 1 0|+ |10\rangle\langle 1 1|$.
Since $d_1^{\dagger}d_2 \leftrightarrow |01\rangle \langle 01|$, we
identify the observable of interest $\sigma_{1,2}$=${\rm Tr}[\rho
d_{1}^{\dagger}d_2]$ by $\sigma_{b,c}$.
In the noninteracting ($U=0$) case, the following equations hold in
in the infinite bias limit \cite{Gurvitz}
\bea \dot\sigma_{a,a}&=& -4\Gamma_L \sigma_{a,a} \nonumber\\
&+&2\Gamma_R\left( \sigma_{b,b}
+\sigma_{c,c}+\sigma_{b,c}e^{i\phi/2} +
\sigma_{c,b}e^{-i\phi/2}\right) \nonumber\\
\dot\sigma_{b,b}&=& 2\Gamma_L \sigma_{a,a}
-2(\Gamma_R+\Gamma_L)\sigma_{b,b} + 2\Gamma_R \sigma_{d,d} \nonumber\\
&+& \delta \Gamma^* e^{i\phi/2}\sigma_{b,c} +\delta\Gamma
e^{-i\phi/2}\sigma_{c,b}
\nonumber\\
\dot\sigma_{c,c}&=& 2\Gamma_L \sigma_{a,a}
-2(\Gamma_R+\Gamma_L)\sigma_{c,c} + 2\Gamma_R \sigma_{d,d}
\nonumber\\ &+& \delta\Gamma^*e^{i\phi/2}\sigma_{b,c} +\delta\Gamma
e^{-i\phi/2}\sigma_{c,b}
\nonumber\\
\dot\sigma_{d,d}&=& 2\Gamma_L\left(\sigma_{b,b} +\sigma_{c,c}
-e^{-i\phi/2}\sigma_{b,c} -e^{i\phi/2}\sigma_{c,b} \right)
\nonumber\\ &-&4\Gamma_R\sigma_{d,d}
 \nonumber\\
\dot\sigma_{b,c}&=&2\Gamma_L e^{i\phi/2}\sigma_{a,a}
+\delta\Gamma(\sigma_{b,b}+\sigma_{c,c})e^{-i\phi/2}
\nonumber\\
&-&2\Gamma_R\sigma_{d,d}e^{-i\phi/2} -2(\Gamma_L+\Gamma_R)\sigma_{b,c}.
\label{eq:GU0}
\eea 
Here $\delta \Gamma=(e^{i\phi}\Gamma_L-\Gamma_R)$. The hybridization
strength, independent of the site index $m$, is defined as
$\Gamma_{\alpha}=\pi\sum_k
V_{L,k,m}^2\delta(\epsilon-\epsilon_k)$. The equations are valid
in the infinite bias limit, when $|\mu_{L}-\mu_R|\gg\Gamma$. 
The total probability, to occupy any of the four states, is
unity, $\sum_{j=a,b,c,d}\sigma_{j,j}=1$.  In the steady-state limit
we demand that $d\vec \sigma/dt=0$,
the vector $\vec\sigma$ includes
the matrix elements ${\sigma_{k,j}}$ of Eq. (\ref{eq:GU0}),
and obtain the stationary solution, valid for $\phi\neq 0$,
\bea
\sigma_{b,c}(t\rightarrow \infty)=\frac{i}{2}\sin(\phi/2).
\label{eq:gur1} \eea
This expression holds in the symmetric setup, $\Gamma_L=\Gamma_R$, for 
$\phi\neq 2\pi p$, $p$ is an integer.
One could formally write  $\sigma_{b,c}(t)=|\sigma_{b,c}(t)|e^{i\varphi(t)}$, noting that
$\varphi$  equals $\pm \pi/2$ in the steady-state limit.
This ``phase-localization" behavior was explored in Ref. \cite{Tu}:
The imaginary part of $\sigma_{b,c}$ depends on the
magnetic phase factor, maximal for $\phi=\pi$ with the value $1/2$. The real part is
identically zero. The results of Figs. \ref{Fig1} and \ref{Fig2}
demonstrate the corresponding behavior at finite bias. There, the real part
is finite, yet small, approaching a fixed value. The imaginary part
slightly deviates from the prediction of Eq. (\ref{eq:gur1}) due to
the finite bias used. One could also get hold of the characteristic rates
from the dynamical equation, by diagonalizing the matrix $M$ in 
$d\vec{\sigma}/dt=M\vec\sigma$. 
We gather five rates, with two phase dependent rates,
$\propto[1\pm\cos(\phi/2)]$. For small $\phi$, the smallest rate is
$\propto[1-\cos(\phi/2)]$, in agreement with \cite{Tu}. 
It can be also proved
that in this noninteracting case the steady-state current scales with $\langle I_e\rangle
\propto [1+\cos(\phi)]$ \cite{Gurvitz}.

\begin{figure}[htbp]
\vspace{1mm} \hspace{0mm} {\hbox{\epsfxsize=70mm \epsffile{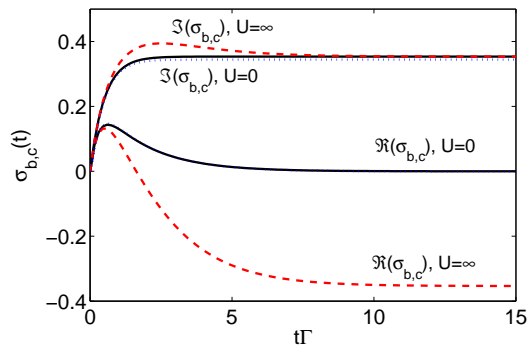}}}
\caption{Master equation analysis: Real and imaginary parts of $\sigma_{b,c}$ at $\phi=\pi/2$,
for $U=0$ and $U=\infty$, obtained by simulating Eq. (\ref{eq:GU0}) and Eq. (\ref{eq:GUinf}),
respectively. Results from INFPI method with $U$=0 are represented by dotted lines,
practically overlapping with master equation curves.}
\label{FigG}
\end{figure}

The dynamics of the coherence, attained from the master equation
(\ref{eq:GU0}), is displayed in Fig. \ref{FigG} for $\phi=\pi/2$. In
the long time limit the real part approaches zero; the imaginary
part reaches $\frac{1}{2}\sin(\pi/4)=0.354$.
INFPI results at zero $U$ are also included in dotted
lines. 
Deviations of INFPI simulations from master equation results can be traced down to
the finite band used within INFPI, in comparison to the infinite-flat
band assumed in the master equation approach. For finite $U$,
we have found that at large bias, $\Delta \mu=2D$, 
INFPI data basically overlaps with the $U=0$ case (not shown) as the system basically stands on the symmetric point.

In the Coulomb blockade regime, for
$U/\Gamma\rightarrow \infty$, a strikingly different behavior is
expected. Starting with the many-body Schr\"odinger equation, one can again
derive the system's equations of motion in the large bias limit while
excluding simultaneous occupancy at both dots, $\sigma_{d,d}=0$. The
following equations of motion are then achieved  \cite{Gurvitz,Gurvitz98}
($\sigma_{a,a}+\sigma_{b,b}+\sigma_{c,c}=1$),
\bea \dot\sigma_{a,a}&=&-4\Gamma_L\sigma_{a,a} \nonumber\\
&+&2\Gamma_R(\sigma_{b,b}+\sigma_{c,c}+\sigma_{b,c}e^{i\phi/2}+\sigma_{c,b}e^{-i\phi/2})
\nonumber\\
\dot\sigma_{b,b}&=&2\Gamma_L\sigma_{a,a} \nonumber\\
&-&2\Gamma_R\sigma_{b,b}-\Gamma_R
(\sigma_{b,c}e^{i\phi/2}+\sigma_{c,b}e^{-i\phi/2}) \nonumber\\
\dot\sigma_{c,c}&=&2\Gamma_L\sigma_{a,a} \nonumber\\
&-&2\Gamma_R\sigma_{c,c}-\Gamma_R(\sigma_{b,c}e^{i\phi/2}+\sigma_{c,b}e^{-i\phi/2}) \nonumber\\
\dot\sigma_{b,c}&=&2\Gamma_Le^{i\phi/2}\sigma_{a,a}
\nonumber\\
&-&\Gamma_Re^{-i\phi/2}(\sigma_{b,b}+\sigma_{c,c})
-2\Gamma_R\sigma_{b,c}. \label{eq:GUinf} \eea
For a spatially symmetric junction, $\Gamma_L=\Gamma_R$, the
steady-state solution for the coherence is
\bea \sigma_{b,c}(t\rightarrow \infty)&=&-\frac{1}{2}e^{-i\phi/2}
\nonumber\\
&=& -\frac{1}{2}\cos(\phi/2)
+\frac{i}{2}\sin(\phi/2).
\label{eq:gur2}
\eea
While the imaginary part predicted is identical to the $U=0$ case,
see Eq. (\ref{eq:gur1}), the real part is finite and phase
dependent. The Coulomb blockade dynamics is presented in Fig.
\ref{FigG} (dashed lines). We find that the imaginary part is
weakly sensitive to the onset of $U$. In contrast, the real part
significantly deviates from the $U=0$ case already at $\Gamma t\sim
1$. 
By analyzing the eigenvalues of the rate matrix (\ref{eq:GUinf}),
we note that phase dependent relaxation rates in the Coulomb blockade regime are the
same as for noninteracting electrons, see also Fig. \ref{FigG}.
It would be interesting to explore this evolution within the INFPI
approach. However, as we are currently limited to $U/\Gamma\leq 4$ values, this
would require an algorithmic improvement of the INFPI technique. We believe that
such an extension could be achieved since the Coulomb blockade case should 
converge faster than the intermediate $U$ limit \cite{RabaniGuy}. This
issue will be tackled in our future work.
A related switching behavior was observed in Ref. \cite{Gurvitz}, 
where the current, finite in the noninteracting case
for $\phi\neq \pi$, vanishes in the Coulomb blockade regime for any phase satisfying $\phi\neq 2 \pi p$.

\section{Summary}

The intrinsic coherence dynamics in a double quantum dot AB
interferometer, away from the symmetric point, was simulated using
an exact numerical technique. At finite interactions,
$U/\Gamma\leq 4$, at low or high temperatures, we have found
that the coherence evolves similarly to the $U=0$
case, showing related characteristic timescales and long time
values. Specifically, we found that for $\phi=\pi/2$ the real part
of $\sigma_{1,2}$ approaches a small number (zero at the symmetric point),
while the imaginary part is larger,
$\sim 0.35$. On the other hand, a master equation treatment in the
Coulomb blockade regime predicts a significantly different behavior:
The magnitude of $\Re \sigma_{1,2}$ and $\Im
\sigma_{1,2}$ should be the same, $\sqrt 2/4$, for the $\phi=\pi/2$ phase factor.

Future work will be devoted to the study of related models,
including the spin degree of freedom at each dot. This model should
demonstrate a decoherence process due to the intrinsic spin-flipping
dephasing effect \cite{GefenPRB}. Other topics of interest are
algorithmic improvement of the INFPI technique, to allow for the
study of the Coulomb blockade regime. One could also add a
local degree of freedom on one of the interferometer arms, e.g.,
a quantum point contact \cite{Aleiner} or a vibrational mode \cite{Thoss},  and observe the
time evolution of the interference pattern in this "which path" experiment.

\acknowledgments DS acknowledges support from NSERC discovery
grant. The research of SB was supported by an Early Research Award
of DS. Fruitful discussions with M. Bandyopadhyay are acknowledged.



\begin{thebibliography}{9}

\bibitem{Rev1}
G. Hackenbroich, Phys. Rep. {\bf343}, 463 (2001), and references therein.

\bibitem{Heilb}
A. Yacoby, M. Heiblum, D. Mahalu, and H. Shtrikman, Phys. Rev. Lett. {\bf 74}, 4047 (1995);
R. Schuster, E. Buks, M. Heiblum, D. Mahalu, V. Umansky, and
H. Shtrikman, Nature {\bf 385}, 417 (1997);
M. Avinun-Kalish, M. Heiblum, O. Zarchin, D. Mahalu, and V.
Umansky, Nature {\bf 436}, 529 (2005).

\bibitem{2QDE}
A. W. Holleitner, C. E. Decker, H. Quin, K. Eberl, and
R. H. Blick, Phys. Rev. Lett. {\bf 87}, 256802 (2001).

\bibitem{Sigrist}
M. Sigrist,T. Ihn, K. Ensslin, D. Loss, M. Reinwald, and W. Wegscheider,
Phys. Rev. Lett. {\bf 96}, 036804 (2006).

\bibitem{GefenPRB}
J. K\"onig and Y. Gefen,
Phys. Rev. Lett. {\bf 86}, 3855 (2001);
Phys. Rev. B {\bf 65}, 045316 (2002).

\bibitem{Gurvitz}
F. Li, X.-Q. Li, W.-M. Zhang, and S. A. Gurvitz,
Euro. Phys. Lett. {\bf 88}, 37001 (2009).

\bibitem{Kubo}
Y. Tokura, H. Nakano, and T. Kubo,
New J.  Phys. {\bf 9}, 113 (2007).

\bibitem{Col}
Y.-S. Liu, H. Chen, and X.-F. Yang,
J. Phys.: Condens. Matt. {\bf 19}, 246201  (2007).


\bibitem{Golosov}
D. I. Golosov and Y. Gefen,
New J. Phys. {\bf 9}, 120 (2007).



\bibitem{Kondo0}
W. Hofstetter, J. K\"onig, and H. Schoeller, Phys. Rev. Lett. {\bf 87}, 156803 (2001).

\bibitem{Boese}
D. Boese, W. Hofstetter, and H. Schoeller, Phys. Rev. B {\bf 66}, 125315 (2002).

\bibitem{KondoHG}
Q.-F. Sun and H. Guo,
Phys. Rev. B {\bf 66},  155308 (2002).

\bibitem{Kondo}
J. Malecki and I. Affleck,
Phys. Rev. B {\bf 82}, 165426 (2010).



\bibitem{Hod}
O. Hod, E. Rabani, and R. Baer, Acc. Chem. Res. {\bf 39}, 109 (2006).

\bibitem{Dhurba}
D. Rai, O. Hod, and A. Nitzan,
J. Phys. Chem. Lett. {\bf 2}, 2118 (2011).


\bibitem{Konig-noint}
B. Kubala and J. K\"onig, Phys. Rev. B {\bf 65}, 245301 (2002).

\bibitem{Green1}
D. Sztenkiel and R. Swirkowicz,
J. Phys.: Condens. Matter {\bf 19} 386224 (2007).

\bibitem{Green2}
Y.-S. Liu, H. Chen, and X.-F. Yang,
J. Phys.: Condens. Matter {\bf 19}  246201 (2007).

\bibitem{Ora}
V. Kashcheyevs, A. Schiller, A. Aharony, and O. Entin-Wohlman,
Phys. Rev. B {\bf 75}, 115313 (2007).


\bibitem{Tu}
M. W.-Y. Tu, W.-M. Zhang, J. Jin, 
Phys. Rev. B {\bf 83}, 115318 (2011).

\bibitem{Tulong}
J. S. Jin, M. W. Y. Tu, 
W. M. Zhang, and Y. J. Yan, New J. Phys.
{\bf 12}, 083013 (2010).

\bibitem{Wacker}
H. A. Nilsson, {\it et al.}, Phys. Rev. Lett. {\bf 104}, 186804 (2010);
O. Karlstr\"om, J. N. Pedersen, P. Samuelsson, A. Wacker,
Phys. Rev. B {\bf 83}, 205412 (2011).

\bibitem{IF1}
D. Segal, A. J. Millis, and D. R. Reichman,
Phys. Rev. B  {\bf 82}, 205323 (2010).

\bibitem{IF2}
D. Segal, A. J. Millis, and D. R. Reichman,
Chem. Phys. Phys. Chem {\bf 13}, 14378 (2011).

\bibitem{NRGBulla}
R. Bulla, T. A. Costi, and T. Pruschke, 
Rev. Mod. Phys. {\bf 80}, 395 (2008).

\bibitem{Anders}
F. B. Anders and A. Schiller, Phys. Rev. Lett. {\bf 95}, 196801 (2005);
Phys. Rev. B {\bf 74}, 245113 (2006);
F. B. Anders, Phys. Rev. Lett. {\bf 101}, 066804 (2008).

\bibitem{Phillip}
E. Gull, A. J. Millis, A. I. Lichtenstein, A. N. Rubtsov, M. Troyer, and P. Werner,
Rev. Mod. Phys. {\bf 83}, 349 (2011).

\bibitem{egger}
S. Weiss, J. Eckel, M. Thorwart, and R. Egger,
Phys. Rev. B {\bf 77}, 195316 (2008);
J. Eckel, F. Heidrich-Meisner, S. G. Jakobs, M. Thorwart, M. Pletyukhov, and R. Egger,
New J. Phys.  {\bf 12}, 043042 (2010).

\bibitem{Anderson}
A. C. Hewson, {\it The Kondo Problem to Heavy Fermions}, (Cambridge
University Press, Cambridge, England, 1993).

\bibitem{Sindel}
M. Sindel, A. Silva, Y. Oreg, and J. von Delft,
Phys. Rev. B {\bf 72}, 125316 (2005).


\bibitem{Nitzan}
M. Galperin, M. A. Ratner, and A. Nitzan,
J. Phys.: Condens. Matter {\bf 19}, 103201 (2007).


\bibitem{Meden}
V. Meden and F. Marquardt,
Phys. Rev. Lett. {\bf 96}, 146801 (2006).

\bibitem{Karrasch}
C. Karrasch, T. Hecht, A. Weichselbaum, J. von Delft, Y.  Oreg, and V. Meden,
New J. Phys. {\bf 9}, 123 (2007).

\bibitem{Hubb-Strat}
J. E. Hirsch, Phys. Rev. B {\bf 28}, 4059 (1983).

\bibitem{Klich}
I. Klich, in "Quantum Noise in Mesoscopic Systems", edited by Yu. V. Nazarov and Ya. M. Blanter (Kluwer, 2003).


\bibitem{RabaniGuy}
G. Cohen and E. Rabani, Phys. Rev. B {\bf 84}, 075150  (2011).

\bibitem{Gurvitz98}
S. A. Gurvitz and Ya. S. Prager, Phys. Rev. B {\bf 53}, 15932 (1996);
S. A. Gurvitz, Phys. Rev. B {\bf 57}, 6602 (1998);
S. A. Gurvitz, D. Mozyrsky, and G. P. Berman,
Phys. Rev. B {\bf 72}, 205341 (2005).

\bibitem{Aleiner}
I. L. Aleiner, N. S. Wingreen, and Y. Meir, 
Phys. Rev. Lett. {\bf 79}, 3740 (1997)

\bibitem{Thoss}
R. H\"artle, C. Benesch, and M. Thoss,
Phys. Rev. Lett. {\bf102}, 146801 (2009).




\end{thebibliography}
\end{document}